\documentclass[12pt]{article}
\usepackage{amssymb,amsmath,graphicx}

\newcommand{\Leff}{\ensuremath{V_{\mathrm{eff}}}}

\newcommand{\Tumn}{\ensuremath{T^{\mu\nu}}}

\newcommand{\gumn}{\ensuremath{g^{\mu\nu}}}

\newcommand{\ket}[1]{|{#1}\rangle}
\begin{document}
%

\begin{center}
{\Large \bf Vacuum Structure and Dark Energy\footnote{
Recognized for Honorable Mention in the Gravity Research Foundation
2010 Awards for Essays on Gravitation, to appear in Int. J. Mod. Phys. D}}\\[0.5cm]
  
{Lance Labun and Johann Rafelski}\\[0.2cm]
{\it Department of Physics, The University of Arizona, Tucson, AZ 85721, USA}\\
\end{center}
\begin{abstract}  

We consider that the universe is trapped in an excited vacuum state and the
resulting excitation energy provides  the observed dark energy.  We explore the
conditions under which this situation can arise from physics already known. 
Considering the example of how macroscopic QED fields alter the vacuum
structure, we find that the energy scale 1 meV --- 1 eV is particularly
interesting. We discuss how dark energy of this form is accessible to laboratory
experiments.

\end{abstract}

An excited quantum vacuum state known as the quark-gluon plasma predominated in the early universe when temperatures exceeded $T_h\sim  160$\,MeV$=1.8\times 10^{12}$\,K  and the age of the Universe was less than $25\mu$s. In the expansion and cooling the quark-filled Universe transformed below $T_h$ into the current matter state with nearly equal  abundances of hadrons and their antiparticles. Drops of  excited quark-deconfined state can be trapped inside hadrons as depicted first by the MIT quark-bag model~\cite{Donoghue:1992dd}. In the subsequent evolution, matter and antimatter annihilated, neutrinos froze out, primordial nucleosynthesis took place and ultimately CMB radiation decoupled---all processes well-known in principle.  We consider here the possibility that the universe relaxed into one of the many nearly-degenerate metastable vacuum states arising within {\it e.g.} the framework of strong interactions, QCD.  This first exploration is carried out in the more manageable environment of QED in the presence of macroscopic applied fields, and our finding determines the scales and is suggestive of beyond the standard model physics.

Since the onset of interest in the deconfined quark-gluon plasma phase, it has been recognized that the early Universe would have as a significant component in the energy balance the cosmological constant-type term usually referred to as $B$ due to the {\bf b}ag model~\cite{Rafelski:2003zz}: a homogeneous distribution of vacuum energy.  The inward pressure of the fluctuations arises since the excited vacuum is necessarily unstable, ready to collapse---the excited vacuum is `pulling' inward on the surrounding True Vacuum $\ket{\rm TV}$, with total collapse prevented by conservation of quantum charges such as baryon number and/or electrical charge.  At the time quarks roamed free in the Universe, the effective dark energy ($=B\simeq (170\,\mathrm{MeV})^4$) was larger by 43 orders of magnitude than it is today.

While the QCD vacuum state is unique, it is `surrounded' on the scale of QCD by a practical continuum of degenerate states, related to the question how strong interactions preserve the CP symmetry~\cite{Ramond:1999vh}.  This is an unresolved issue, and for our purposes it suffices to observe that the Universe may relax from the highly excited deconfined state to one of the many available metastable QCD vacuum states that are (measured on QCD scale) very near to the true ground state.  As the Universe continues to evolve, transitions between these states can occur spontaneously, especially at higher temperatures releasing energy~\cite{Rafelski:2009di}, adding heat to the the thermal bath, and thus improving the homogeneity of the CMB radiation.  In the absence of physics strongly breaking the degeneracy, the Universe at a sufficiently low ambient temperature may freeze into a low-lying excited state which has a lifespan greater than  cosmological times.

The QCD vacuum structure thus offers a framework near to the known physics that leads naturally to consideration of dark energy and the observed cosmological acceleration~\cite{Frieman:2008sn,Silvestri:2009hh}.  However, we must remember that a full understanding of the QCD vacuum remains elusive.  For this reason and in order to gain more quantitative grasp on  the hypothesis that the Universe is in an excited vacuum state, the present investigation is carried out in the more manageable environment of QED in the presence of macroscopic applied fields. 

We explore the scales relevant if dark energy arises as the energy of a vacuum excited by a universe-spanning constant electric and/or magnetic field.  In this situation, we find a simulacrum to dark energy~\cite{Labun:2008qq} and we can study the metastability and energy content of the excited vacuum.  The novel element explored here is the magnitude of mass scale required to explain dark energy in this context.

The vacuum filled with electromagnetic field is a metastable state because the conversion of field energy into particle-antiparticle pairs is possible.  The process is very slow except in very strong fields
\begin{equation}
E=\beta^{-1} \frac{m^2}{e},\quad \beta^{-1} \to 1
\end{equation}
introducing $\beta^{-1}$ as an inverse Hawking-Unruh temperature parameter.  $2m$ is the energy stability gap, $2/m$ the width of the tunneling barrier, and $e$ the coupling charge.  $\beta^{-1}\simeq 0.1$ suffices to induce just one decay event in a macroscopic volume, while full collapse of the vacuum ensues for $\beta^{-1}\to 1$~\cite{Labun:2008re}. 

As dark energy is a long wavelength phenomenon, we study the long-wavelength limit of QED in prescribed external fields of arbitrary strength with small values of the charge $e^2/4\pi^2=\alpha/\pi$. The field is constant and homogeneous, filling our model universe and hence the relevant object to study is the effective Euler-Heisenberg-Schwinger QED action~\cite{EKHSchwinger:1951nm}.  It can be written in the form 
\begin{equation}\label{Veff}
\Leff \equiv -\mathcal{S} + m^4 f_{\rm eff}\left(\frac{\mathcal S}{m^4},\frac{\mathcal P}{m^4}\right);
\end{equation}
where $f_{\rm eff}$, found in textbooks, is a highly nonlinear function of its arguments $\mathcal{S}=\frac{1}{4}F^{\alpha\beta}F_{\alpha\beta}$ and $\mathcal{P}=\frac{1}{4}\,^*\!F^{\alpha\beta}F_{\alpha\beta}$.  Well-understood charge renormalization terms also enter Eq.\,\eqref{Veff} and are omitted here. The total energy-momentum tensor  
\begin{align} \label{SchTmn}
\Tumn &=
   \left(-\frac{\partial \Leff}{\partial \mathcal{S}}\right)
   (\gumn \mathcal{S}- F^{\mu\lambda}F^{\nu}_{\phantom{\nu}\lambda})
-\gumn \left(\Leff-\mathcal{S}\frac{\partial \Leff}{\partial\mathcal{S}}
     -\mathcal{P}\frac{\partial \Leff}{\partial \mathcal{P}} \right)
\end{align}
exhibits the relation of the energy-momentum trace to the scale $m$
\begin{equation}\label{dVdm}
T^{\mu}_{\mu}= 
-4\left(\Leff-\mathcal{S}\frac{\partial \Leff}{\partial \mathcal{S}}
           -\mathcal{P}\frac{\partial \Leff}{\partial \mathcal{P}}\right)
=-m\frac{d{f_{\rm eff}}}{dm} 
\end{equation}

The Euler-Heisenberg-Schwinger calculation exhibits how the electromagnetic field is responsible for the `false vacuum' by pushing the electron-positron fluctuations away from $\ket{\rm TV}$.  The augmented energy of these fluctuations implies the noted instability of the with-field vacuum and leads to the dark energy
\begin{equation}\label{darkdef}
\lambda\to \frac{T^{\mu}_{\mu}}{4}. 
\end{equation}
Classical fields generating $T^{\mu}_{\mu}$ and satisfying the long wavelength approximation are created regularly in experiments.  The reason they do not usually provide insight on dark energy, vacuum structure and instability is most clear from the timescale $\tau$ corresponding to the lifespan of with-field vacuum state.  $\tau$ being exponential in $\beta$ generally far exceeds the age of the universe: for a (QED) electric field very strong by laboratory standards, $\sim 10^{10}$ V/m, the associated timescale is $10^{54}$ years.  This field is chosen as our example because it generates an energy-momentum trace comparable to the dark energy density
\begin{equation}\label{lambdaD}
T^{\mu}_{\mu}(E=32.4\,{\rm GV/m}) 
=6.09\times 10^{-10}\,{\rm J/m}^3 
=(2.3\,{\rm meV})^4 
\end{equation}

Setting the scale $m$ at the electron mass (0.5 MeV) is the cause in this example of the discordant result that a field gargantuan in laboratory terms is required to induce a small vacuum energy.  We can ask what scale $m$ admits a field that has a lifetime exceeding the age of the universe and hence can persist up to the present while generating the observed dark energy.  No optimization is attempted here and in particular the coupling is left fixed at the QED $\alpha$, but in the figure, we see that the strength of the field must remain below $\beta^{-1} \simeq .04$ for the excited state to live as long as the universe without decaying.  Up to factors of order unity, $m$ is found in the range 1 meV -- 1 eV  which is suggestive that dark energy is a phenomenon buried in depths of and beyond the standard model physics, related to the above-mentioned QCD vacuum degeneracy and/or neutrinos and their masses~\cite{Bjaelde:2008yd}.  The non-perturbative maximum in the energy-momentum trace at $\beta^{-1}\simeq 5.6$ and subsequent change in sign are manifestations of the instability of the electric field, resulting from the rapidly increasing imaginary part~\cite{Labun:2008qq}.

\begin{figure*}
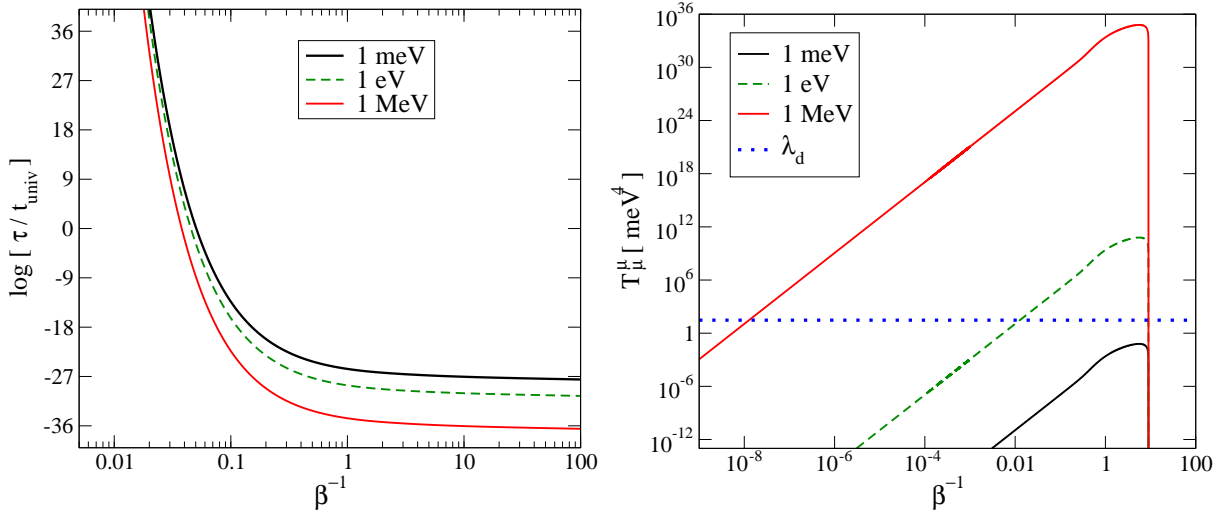

\includegraphics[width=.48\textwidth]{LabunRafelski_1a.eps}
\includegraphics[width=.49\textwidth]{LabunRafelski_1b.eps}
\caption{At left, the lifetimes of electric fields $E = \beta^{-1}m^2/e$ in terms of the age of the universe $t_{\rm univ}$ for various values of the mass scale.  Note that for $m=1$ meV, a field of .25 V/m ($\beta = 0.05$) has a lifetime equal to the age of universe, and any smaller field will survive into the present era without decaying.  At right, the field induced energy-momentum trace with the dark energy density at the horizontal line for comparison.
\label{thefig}}
\end{figure*}

We can hope to manipulate soon this field-induced simulacrum of the dark energy. High intensity pulsed laser experiments will achieve electromagnetic fields closer to the critical scale $E\to m^2/e$, that is $\beta^{-1} \to 1$, even when $m=m_e$ and hence allow the study $T^{\mu}_{\mu}$ and the vacuum structure. The scale we invoke $0.001 <m<1$\,eV  is already fully accessible in the laboratory.  With wavelengths in the optical or infrared, the long wavelength condition holds, and many phenomena associated with the field and its lifetime are expected~\cite{Mourou:2006zz}.  Equally and even more interesting, at critical field strengths we may also expect novel effects in the acceleration of an electron in such a field, because
\begin{equation}\label{unit-accel}
m\frac{dp}{dt} = eE\to m^2, \qquad\mathrm{i.e.}~~ \frac{dp}{d(mt)} = 1, 
\end{equation}
that is `unit' acceleration in natural units, with $t$   measured in units of $m$.  Just as for strong gravity, new physics is expected, not least because radiation reaction is no longer negligible but rather a dominant effect.

We see that the critical fields where the vacuum is unstable are directly related to the yet unexplored strong acceleration phenomena in situations corresponding to the most singular gravity.  Eq.\,\eqref{unit-accel} also recalls the connection of vacuum instability to Hawking-Unruh radiation, which is characterized by the Hawking-Unruh temperature
\begin{equation}
T_{\rm HU} = \frac{a}{2\pi},
 \quad\mathrm{where}~~ a=\frac{eE}{m}=m\beta^{-1}
\end{equation}
is the acceleration of an electron in an electric field $E$. 

One recognizes that fields causing acceleration at the scale of unity connect to an unstable effective action describable  in the framework of the Hawking-Unruh temperature, though some technical nuances still need to be resolved~\cite{PauchyHwang:2009rz}. High intensity pulsed laser-electron collisions in which a relativistic charged particle is stopped within one wavelength generate critical acceleration and thus can test the limit of the current concept of interactions. Other experimental approaches  include LHC heavy ion reactions and in the foreseeable future laser pulse-on-laser pulse collisions.  By probing consequences of the force scale $m^2/e$, these experiments also access the nature of the excited vacuum.  

We conclude that resolving vacuum structure in strong fields is  a natural and well-defined experimental and theoretical context with results that may bear on the mystery of dark energy. Its consequences are available to laboratory study by exploring strong fields and the effects of the induced energy-momentum trace $T^{\mu}_{\mu}$ and critical acceleration. The ultimate test for an excited vacuum state generating the dark energy is the observation of a release of the excitation energy associated with the vacuum transition.  This process has recently been investigated for the de Sitter geometry, which is the gravitational analog of the pair-creating electric field~\cite{Polyakov:2009nq}.  In view of their ability to affect vacuum structure, it can be hoped that strong applied QED fields could catalyze decay of the `false' vacuum. 

\subsubsection*{Acknowledgments}
This work was supported by a grant from the U.S. Department of Energy  DE-FG02-04ER41318.

\clearpage

\vspace{-0.3cm}

\begin{thebibliography}{17}

\bibitem{Donoghue:1992dd}
  J.~F.~Donoghue, E.~Golowich and B.~R.~Holstein,
  {\it Dynamics of the Standard Model},
 (University Press, Cambridge, 1992).

\bibitem{Rafelski:2003zz}
  J.~Kapusta, B.~M\"uller and J.~Rafelski,
  {\it Quark-Gluon Plasma: Theoretical Foundations}  
  (Elsevier, Amsterdam, 2003).

\bibitem{Ramond:1999vh}
  P.~Ramond,
  {\it Journeys Beyond The Standard Model}
  (Perseus Books, Reading, 1999).

\bibitem{Rafelski:2009di}
  J.~Rafelski, L.~Labun, Y.~Hadad and P.~Chen,
  in {\it Tenth Workshop on
  Non-Perturbative Quantum Chromodynamics}
  (SLAC eConf Proceedings, 2009) C0906083:
  http://www.slac.stanford.edu/econf/C0906083/.

\bibitem{Frieman:2008sn}
  J.~Frieman, M.~Turner and D.~Huterer,
  Ann.\ Rev.\ Astron.\ Astrophys.\  {\bf 46} (2008) 385. 

\bibitem{Silvestri:2009hh}
  A.~Silvestri and M.~Trodden,
  Rept.\ Prog.\ Phys.\  {\bf 72} (2009) 096901.
  
\bibitem{Labun:2008qq}
  L.~Labun and J.~Rafelski,
  Phys.\ Rev.\ D {\bf 81} (2010) 065026.  


\bibitem{Labun:2008re}
  L.~Labun and J.~Rafelski,
  Phys.\ Rev.\  D {\bf 79} (2009) 057901.

\bibitem{EKHSchwinger:1951nm}  
  H.~Euler and B.~Kockel, Naturwiss. {\bf 23} (1935) 246;
  H.~Euler, Ann.\ Physik.\ (Berlin) {\bf 26} (1936) 398; 
  W.~Heisenberg and H.~Euler,
  Z.\ Phys.\  {\bf 98}  (1936) 714, 
  for translation see [arXiv:physics/0605038];
  J.~S.~Schwinger,
  Phys.\ Rev.\  {\bf 82} (1951) 664.

\bibitem{Bjaelde:2008yd}
  O.~E.~Bj\ae lde and S.~Hannestad,
  Phys.\ Rev.\  D {\bf 81} (2010)  063001.

\bibitem{Mourou:2006zz}
  G.~A.~Mourou, T.~Tajima and S.~V.~Bulanov,
  Rev.\ Mod.\ Phys.\  {\bf 78} (2006) 309.

\bibitem{PauchyHwang:2009rz}
  W.~Y.~Pauchy Hwang and S.~P.~Kim,
  Phys.\ Rev.\  D {\bf 80} (2009) 065004.
  
\bibitem{Polyakov:2009nq}
  A.~M.~Polyakov,
  Nucl.\ Phys.\  B {\bf 834} (2010) 316.

\end{thebibliography}
\end{document}